\documentclass[twocolumn,aps,pra,showpacs,superscriptaddress,floatfix]{revtex4}

\usepackage{graphicx}
\usepackage{times}
\usepackage{nicefrac}
\usepackage{amsmath}
\usepackage{amsfonts}
\usepackage{amssymb}
\usepackage{amsthm}
\usepackage{epsf}
\usepackage{bm}
\usepackage{bbm}

\usepackage{dcolumn}
\newcolumntype{.}{D{x}{}{-1}}

\begin{document}

\newcommand{\vare}{\varepsilon}

\newcommand{\pr}{^{\prime}}

\newcommand{\bfx}{{\bf {x}}}
\newcommand{\bfy}{{\bf {y}}}
\newcommand{\bfz}{{\bf {z}}}
\newcommand{\bfp}{{\bf {p}}}
\newcommand{\bfq}{{\bf {q}}}

\newcommand{\hx}{\hat{\bfx}}

\newcommand{\beq}{\begin{equation}}
\newcommand{\eeq }{\end{equation}}
\newcommand{\beqn}{\begin{eqnarray}}
\newcommand{\eeqn }{\end{eqnarray}}
\newcommand{\ba}{\begin{array}}
\newcommand{\ea}{\end{array}}
\newcommand{\balpha}{\bm{\alpha}}
\newcommand{\Za}{Z\alpha}

\newcommand{\lbr}{\langle}
\newcommand{\rbr}{\rangle}

\newcommand{\Dmatrix}[4]{
        \left(
        \begin{array}{cc}
        #1  & #2   \\
        #3  & #4   \\
        \end{array}
        \right)
        }
\newcommand{\im}{{\rm i}}

\title{One-loop self-energy correction in a strong binding field}

\author{Vladimir A. Yerokhin}
\affiliation{Department of Physics, St.~Petersburg State
University, Oulianovskaya 1, Petrodvorets, St.~Petersburg 198504,
Russia}
\affiliation{Center for Advanced Studies, St.~Petersburg
State Polytechnical University, Polytekhnicheskaya 29,
St.~Petersburg 195251, Russia}

\author{Krzysztof Pachucki}
\affiliation{Institute of Theoretical Physics, Warsaw University,
ul.~Ho\.{z}a 69, 00--681 Warsaw, Poland}

\author{Vladimir M. Shabaev}
\affiliation{Department of Physics, St.~Petersburg State
University, Oulianovskaya 1, Petrodvorets, St.~Petersburg 198504,
Russia}

\begin{abstract}
A new scheme for the numerical evaluation of the one-loop
self-energy correction to all orders in $\Za$ is presented. The
scheme proposed inherits the attractive features of the standard
potential-expansion method but yields a partial-wave expansion
that converges more rapidly than in the other methods reported in
the literature.

\end{abstract}

\maketitle

\section{Introduction}

Calculations of the one-loop self-energy correction to all orders
in the parameter $\Za$ ($Z$ is the nuclear charge number and
$\alpha$ is the fine-structure constant) have a long history. The
first correct evaluation of this correction was performed for
several high-$Z$ ions by Desiderio and Johnson \cite{desiderio:71}
using the method proposed by Brown, Langer, and Schaefer
\cite{brown:59:procI}. Another, much more accurate and powerful
method was developed by Mohr \cite{mohr:74:a}, who carried out
high-precision calculations of this correction in a wide range of
$Z$ for the ground and the first excited states of H-like ions
\cite{mohr:74:b,mohr:82,mohr:92:b}. Various extensions of this
method provided highly accurate evaluations of the self-energy
correction for higher excited states \cite{mohr:92:a,bigot:01:se},
for the extended-nucleus Coulomb potential \cite{mohr:93:prl}, and
for very small nuclear charge numbers
\cite{jentschura:99:prl,jentschura:01:pra,jentschura:04:se}.
Indelicato and Mohr \cite{indelicato:92:se,indelicato:98}
presented an important modification of the method, in which
renormalization is performed completely in coordinate space.

A different method for evaluation of the self-energy correction,
which can  be conventionally termed as the potential-expansion
method, was introduced by Snyderman and Blundell
\cite{snyderman:91,blundell:91:se,blundell:92}. Various numerical
schemes based on this method were presented by other groups
\cite{cheng:93,yerokhin:99:pra}.

There are also other methods developed for evaluation of the
self-energy correction which have been less widely used so far. A
noncovariant method of the so-called partial-wave renormalization
was developed by Persson, Lindgren, and Salomonson
\cite{persson:93:ps} and by Quiney and Grant
\cite{quiney:93,quiney:94}. Another method proposed by Labzowsky
and Goidenko \cite{labzowsky:97:mcm} is based on the multiple
commutator expansion of the general expressions.

Closely related to the self-energy is the other dominant QED
effect, the vacuum-polarization. The first evaluations of this
correction to all orders in $\Za$ were performed by Soff and Mohr
\cite{soff:88:vp} and by Manakov, Nekipelov, and Fainstein
\cite{manakov:89:zhetp}. More accurate calculations of the
vacuum-polarization correction were carried out later by other
groups \cite{persson:93:vp,sapirstein:03:vp}.

Evaluation of the self-energy correction for a tightly bound
electron is nontrivial, to a large extent, due to the fact that
this correction involves the Dirac-Coulomb Green function that is
not presently known in the closed analytical form (contrary to the
nonrelativistic Coulomb Green function). Consequently, the
self-energy correction is expressed as an infinite expansion over
the angular momentum of the virtual photon (or, equivalently, the
total angular momentum of the virtual electron states $j =
|\kappa|-1/2$, where $\kappa$ is the relativistic angular-momentum
parameter of the Dirac equation). This expansion (further referred
to as the partial-wave expansion) greatly complicates calculations
of the self-energy corrections.

In the method by Mohr \cite{mohr:74:a}, the summation of the
partial-wave expansion was performed numerically before
integrations over radial coordinates. A large number of terms
included into the summation ($\sim 10^4$) and usage of the
quadruple arithmetics ensured a high accuracy of the numerical
results obtained but made the computation rather time consuming.
In the extension of this method by Jentschura {\em et al.}
\cite{jentschura:99:prl,jentschura:01:pra,jentschura:04:se},
several millions of expansion terms included into computation were
reported, which became possible due to an elaborate
convergence-acceleration technique developed by the authors and an
extensive usage of modern parallel computer systems.

On the contrary, calculations based on the potential-expansion
method \cite{blundell:91:se,blundell:92,cheng:93,yerokhin:99:pra}
are usually performed with much smaller numbers of partial-wave
expansion terms actually included into the computation ($\sim
15-40$). This is achieved ({\em i}) by employing a more complete
set of renormalization terms that are calculated separately in a
closed form, ({\em ii}) by performing the radial integrations
before the partial-wave summation (for the discussion of how this
influences the convergence rate see Eqs.~(1), (2) of
Ref.~\cite{indelicato:98} and the related text there), and ({\em
iii}) by using extrapolation to estimate the contribution of the
tail of the expansion. The price to pay is a more complex
structure of the subtraction terms (especially, in coordinate
space) and the necessity to keep the accuracy of numerical
integrations well under control for each partial-wave term, in
order to provide a reasonable extrapolation for the tail of the
expansion. Still, the method is computationally very cheap and can
be directly generalized for calculations of higher-order QED
diagrams, where the self-energy loop enters as a subgraph. These
advantages have determined the fact that most calculations of
higher-order self-energy corrections have been performed by
extensions of the potential-expansion method up to now.

The one-loop self-energy correction is traditionally represented
in terms of the dimensionless function $F(\Za)$, which is
connected to the energy shift (in units $\hbar = c = m = 1$) by
\begin{equation}\label{0a}
    \Delta E = \frac{\alpha}{\pi}\,\frac{(\Za)^4}{n^3}\,F(\Za)\,,
\end{equation}
where $n$ is the principal quantum number. Practical calculations
performed within the potential-expansion method in the Feynman
gauge show that the general behavior of individual partial-wave
expansion contributions to the function $F(\Za)$ roughly follows
the dependence
\begin{equation}\label{0b}
    F_{|\kappa|} \approx \frac{n^3}{10\, (\Za)^2\, |\kappa|^3}\,.
\end{equation}
This makes clear that, while the nominal rate of convergence of
the partial-wave expansion is always close to $|\kappa|^{-3}$ in
this method, the actual convergence is governed by the parameter
$n^3/(\Za)^2$, whose numerical value can be rather large for
excited states and small nuclear-charge numbers. Taking into
account that the extension of the partial-wave summation beyond
the typical limit of $|\kappa|=30-40$ leads to serious technical
problems within the numerical scheme employed, we conclude that
the parameter $n^3/(\Za)^2$ defines the region of the practical
applicability of the potential-expansion method.

Similar situation persists in calculations of self-energy
corrections to higher orders of perturbation theory. In such
calculations, the convergence of the partial-wave expansion also
worsens with decrease of $Z$ and increase of $n$. In particular, a
slow convergence of this expansion turned out to be the factor
limiting the accuracy in evaluations of the self-energy correction
to the $1s$ and $2s$ hyperfine splitting in low-$Z$ ions
\cite{blundell:97:prl,sunnergren:98:pra,yerokhin:01:hfs}. This
convergence also posed serious problems in calculations of the
self-energy correction to the bound-electron $g$ factor in light
H-like ions
\cite{blundell:97:pra,persson:97:g,beier:00:pra,yerokhin:02:prl}.

The convergence rate of the partial-wave expansion becomes most
crucial in the case of two-loop self-energy corrections, for which
the summation should be performed over two independent expansion
parameters, both of which are unbound
\cite{mallampalli:98:pra,yerokhin:03:prl}. A calculation of the
two-loop self-energy correction for very low nuclear charge
numbers (and, specifically, for hydrogen) is a challenging
problem, which apparently cannot be solved within a
straightforward generalization of the potential-expansion method.
(The present status of calculations of the two-loop self-energy
correction can be found in Ref.~\cite{yerokhin:05:sese}). One of
the problems to be solved to this end is to find a way to improve
the convergence properties of the partial-wave expansion.

The goal of the present investigation is to formulate a scheme for
evaluation of the one-loop self-energy correction, which yield the
fastest convergence of the partial-wave expansion among the
methods reported so far in the literature.

%
\section{Formalism}

The energy shift of a bound electron due to the first-order
self-energy correction is given by the real part of the expression
\begin{eqnarray} \label{3}
\Delta E  &=& 2\,\im\, \alpha \int_{-\infty}^{\infty} d\omega \int
d\bfx_1\,
        d \bfx_2\, D^{\mu \nu}(\omega, \bfx_{12})
  \nonumber \\ && \times
        \psi^{\dag}_a(\bfx_1)\, \alpha_{\mu}\,
        G(\vare_a-\omega,\bfx_1,\bfx_2)\, \alpha_{\nu}\,
        \psi_a(\bfx_2)\,
  \nonumber \\ &&
   - \delta m \int d\bfx\, \psi^{\dag}_a(\bfx)\, \beta \, \psi_a(\bfx) \, ,
\end{eqnarray}
where $\alpha_{\mu} = (1,\balpha)$, $\balpha$ and $\beta$ are the
Dirac matrices, $G(\omega,\bfx_1,\bfx_2) = [\omega-{\cal H}(1-\im
0)]^{-1}$, ${\cal H}={\cal H}_0+ V(x)$, ${\cal H}_0 = \balpha
\cdot \bfp + \beta $ is the free Dirac Hamiltonian, $V(x)$ is a
local potential (not necessarily the Coulomb one), and $\delta m$
is the mass counterterm. $D^{\mu\nu}$ is the photon propagator
defined in the Feynman gauge as
\begin{equation} \label{4}
D^{\mu \nu}(\omega, \bfx_{12}) = g^{\mu \nu}\,
        \frac{\exp( \im \sqrt{\omega^2+\im 0}\, x_{12})}
        {4\pi \, x_{12}} \ ,
\end{equation}
where $x_{12} = |\bfx_{12}|= |\bfx_1-\bfx_2|$, and the branch of
the square root is fixed by the condition ${\rm
Im}(\sqrt{\omega^2+\im 0})
> 0$. In Eq.~(\ref{3}) it is assumed that the unrenormalized part
of the expression and the mass counterterm are regularized in a
certain covariant way and that the limit removing the
regularization is taken after the cancellation of the divergent
terms.

Ultraviolet divergencies in Eq.~(\ref{3}) can be conveniently
isolated by separating the first two terms in the expansion of the
bound-electron propagator $G$ in terms of the binding potential
$V$,
\begin{eqnarray} \label{5}
G(E,\bfx_1,\bfx_2) &=&
    G^{(0)}(E,\bfx_1,\bfx_2)+G^{(1)}(E,\bfx_1,\bfx_2)
        \nonumber \\ &&
        {}+G^{(2+)}(E,\bfx_1,\bfx_2)\,,
\end{eqnarray}
where $G^{(0)} = [\omega-{\cal H}_0(1-\im 0)]^{-1}$ is the free
Dirac Green function, $G^{(1)}$ is the first-order expansion term
\begin{equation} \label{6}
G^{(1)}(E,\bfx_1,\bfx_2) = \int d\bfz\, G^{(0)}(E,\bfx_1,\bfz)\,
             V(z)\, G^{(0)}(E,\bfz,\bfx_2)\,,
\end{equation} and $G^{(2+)}$ is the remainder. The three terms in
Eq.~(\ref{5}), after substitution into Eq.~(\ref{3}), lead to the
separation of the self-energy correction into the zero-potential,
one-potential, and many-potential parts \cite{snyderman:91}:
\begin{equation}\label{7}
 \Delta E = \Delta E_{\rm zero}+  \Delta E_{\rm one}+
                        \Delta E_{\rm many}\,,
\end{equation}
with the mass-counterterm part naturally ascribed to the
zero-potential term. Converting the first two terms into momentum
space and cancelling the ultraviolet divergences, one obtains:
\begin{equation}\label{7a}
    \Delta E_{\rm zero}= \int \frac{d\bfp}{(2\pi)^3} \,\,
        \overline{\psi}_a(\bfp)\,  \Sigma^{(0)}_R(\vare_a,\bfp)\, \psi_a(\bfp)\, ,
\end{equation}
\begin{eqnarray}\label{7b}
    \Delta E_{\rm one} &=&
        \int \frac{d\bfp_1}{(2\pi)^3}\,
         \frac{d\bfp_2}{(2\pi)^3} \,\,
        \overline{\psi}_a(\bfp_1)\,
 \nonumber \\ && \times
        \Gamma^0_R(\vare_a,\bfp_1;\vare_a,\bfp_2)\,
        V(\bfq)\, \psi_a(\bfp_2) \, ,
\end{eqnarray}
where $\bfq = \bfp_1-\bfp_2$, $\overline{\psi}_a(\bfp) =
\psi_a^{\dag}(\bfp)\, \gamma^0$, and $\Sigma^{(0)}_R(p)$ and
$\Gamma^{\mu}_R(p_1,p_2)$ are the renormalized free self-energy
and vertex functions (for their exact definition and calculational
formulas see, {\em e.g.}, Ref.~\cite{yerokhin:99:pra}).

The many-potential term is represented by the following expression
\begin{align} \label{8}
\Delta E_{\rm many}  =& \ 2\,\im\, \alpha \int_{C} d\omega
        \int d\bfx_1\,
        d \bfx_2\, D^{\mu \nu}(\omega, \bfx_{12})\,
  \nonumber \\  & \times
        \psi^{\dag}_a(\bfx_1)\, \alpha_{\mu}\,
        G^{(2+)}(\vare_a-\omega,\bfx_1,\bfx_2)\, \alpha_{\nu}\,
        \psi_a(\bfx_2)\, ,
        \nonumber \\
\end{align}
where $G^{(2+)} = G - G^{(0)}- G^{(1)}$ and the contour $C$ of the
$\omega$ integration does not necessarily go along the real axis
but can be chosen differently in order to simplify the numerical
evaluation of this expression. In our approach, we employ the
contour $C_{LH}$ that consists of the low-energy part ($C_L$) and
the high-energy part ($C_H$) and is similar to the one introduced
in our previous work \cite{yerokhin:99:pra}. The low-energy part
of the contour $C_L$ extends from $-\vare_0-\im 0$ to $-\im 0$ on
the lower bank of the branch cut of the photon propagator and from
$+\im 0$ to $\vare_0+\im 0$ on the upper bank of the cut. In order
to avoid appearance of poles of the electron propagator near the
integration contour, each part of $C_L$ is bent into the complex
plane if the calculation is performed for excited states. (The
analytical structure of the integrand and a possible choice of the
contour are discussed in Ref.~\cite{yerokhin:99:pra}.) The
high-energy part of the contour is $C_H = (\vare_0-\im
\infty,\vare_0-\im 0] + [\vare_0+\im 0,\vare_0+\im \infty)$. The
parameter $\vare_0$ separating the low- and the high-energy part
of the contour is chosen to be $\vare_0 = \Za\,\vare_a$ in this
work. (It is assumed that the condition
$\vare_a-\vare_{1s}<\vare_0$ is fulfilled for the states under
consideration, where $\vare_{1s}$ is the ground-state energy.)

Due to a lack of a closed-form representation for the Dirac
Coulomb Green function, the evaluation of the many-potential term
has to be performed by expanding $G$ (and, therefore, $G^{(2+)}$)
into eigenfunctions of the Dirac angular momentum with the
eigenvalue $\kappa$. As discussed in Introduction, the convergence
rate of the resulting partial-wave expansion is of crucial
importance for the numerical evaluation of the self-energy
correction.

Until this moment, our description closely followed the standard
potential-expansion method \cite{snyderman:91}. We would like now
to modify this method in order to achieve a better convergence of
the partial-wave expansion in the many-potential term $\Delta
E_{\rm many}$. To this end, we look for an approximation
$G_a^{(2+)}$ to the function $G^{(2+)}$ that fulfills the
following requirements: ({\em i}) it can be evaluated in a closed
form ({\em i.e.}, without the partial-wave expansion) and ({\em
ii}) the difference $G^{(2+)} - {G}_a^{(2+)}$ inserted into
Eq.~(\ref{8}) yields a rapidly converging partial-wave series.

We start with the expansion of the bound-electron Green function
in terms of the binding potential,
\begin{widetext}
\begin{eqnarray}\label{9}
    G(E,\bfx_1,\bfx_2) &=& G^{(0)}(E,\bfx_1,\bfx_2)
       + \int d\bfz\, G^{(0)}(E,\bfx_1,\bfz)\, V(z)\, G^{(0)}(E,\bfz,\bfx_2)
 \nonumber \\ &&
      +  \int d\bfz_1\, d\bfz_2\, G^{(0)}(E,\bfx_1,\bfz_1)\, V(z_1)\,
               G^{(0)}(E,\bfz_1,\bfz_2)\, V(z_2)\,
                    G^{(0)}(E,\bfz_2,\bfx_2)+ \ldots \,.
\end{eqnarray}
\end{widetext}
It is well known that the dominant contribution to radial
integrals like those that appear in Eq.~(\ref{8}) originates from
the region where the radial arguments are close to each other,
$\bfx_1 \approx \bfx_2$. This region is also responsible for the
part of the partial-wave expansion of the Green function that has
the slowest asymptotic convergence in $1/|\kappa|$
\cite{mohr:74:a}. In this region the commutators of the potential
$V$ with the free Green function $G^{(0)}$ are small and can be
neglected, which corresponds to expanding $V(\bfz)$ in a Taylor
series around $\bfz=\bfx_1$ (or $\bfx_2$) and keeping only the
first term. Commuting $V$ out to the left in Eq.~(\ref{9}) and
repeatedly employing the identity
\begin{align}\label{10}
    \int d\bfz\, G^{(0)}(E,\bfx_1,\bfz)\,
G^{(0)}(E,\bfz,\bfx_2)
   \nonumber \\
    = -\frac{\partial}{\partial E}\, G^{(0)}(E,\bfx_1,\bfx_2)\,,
\end{align}
we obtain the approximation ${G}_a$ to the bound-electron Green
function $G$,
\begin{align}\label{11}
G_a(E,\bfx_1,\bfx_2) =&\ \ G^{(0)}(E,\bfx_1,\bfx_2)
 \nonumber \\ &
       -V(x_1)\, \frac{\partial}{\partial E}\, G^{(0)}(E,\bfx_1,\bfx_2)
 \nonumber \\ &
       + V^2(x_1)\, \frac{\partial^{\,2}}{\partial E^{\,2}}\, G^{(0)}(E,\bfx_1,\bfx_2)
       + \ldots \,.
 \end{align}
This expansion has a form of the Taylor series and can be formally
summed up, yielding
\begin{equation}\label{11a}
    G_a(E,\bfx_1,\bfx_2)
       = G^{(0)}(E+\Omega,\bfx_1,\bfx_2)\,,
\end{equation}
where $\Omega = -V(x_1) = \Za/x_1$. Commuting $V$ out to the right
in Eq.~(\ref{9}), we obtain the same representation for $G_a$ but
with $\Omega = \Za/x_2$.

It should be noted that the idea of commuting the potential $V$
outside in the one-potential term was first proposed by Mohr
\cite{mohr:74:a}, who proved that this procedure does not
influence the asymptotic ultraviolet behavior of this term (we
recall that ultraviolet divergences originate from the region
$\bfx_1 \approx \bfx_2$ in configuration space). Later, it was
also demonstrated \cite{indelicato:92:se,indelicato:98} that all
ultraviolet divergences in the one-loop self-energy correction
could be identified by isolating several first terms of the
power-series expansion of the potential $V$ and the
reference-state wave functions $\psi_a$ around the point
$\bfx_1=\bfx_2$.

Expression (\ref{11a}) yields an approximation for the
bound-electron Green function that has a form of the free Green
function with a shifted energy argument. Taking into account that
the free Green function is known in a closed form \cite{mohr:74:a}
\begin{eqnarray}\label{12}
    G^{(0)}(E,\bfx_1,\bfx_2) &=& -\left[
\left(\frac{c}{x_{12}}+\frac1{x_{12}^2}\right)
           \im\, \balpha \cdot \bfx_{12} +\beta +E \right]\,
           \nonumber \\ && \times
                  \frac{{\rm exp}[-c\,x_{12}]}{4\pi x_{12}}\,,
\end{eqnarray}
($c = \sqrt{1-E^2}\,$), we can employ this expression for the
evaluation of $G_a$.

An analogous to Eq.~(\ref{11a}) approximation for the function
$G^{(2+)}$ is obtained by subtracting the first two terms of the
Taylor expansion from $G_a$,
\begin{align}\label{13}
    G_a^{(2+)}(E,\bfx_1,\bfx_2)  =&\
G^{(0)}(E+\Omega,\bfx_1,\bfx_2)
   - G^{(0)}(E,\bfx_1,\bfx_2)
           \nonumber \\ &
   -         \Omega\, \frac{\partial}{\partial E}\, G^{(0)}(E,\bfx_1,\bfx_2)\,.
\end{align}

According to the derivation, the functions $G_a(E,\bfx_1,\bfx_2)$
and $G_a^{(2+)}(E,\bfx_1,\bfx_2)$ approximate, correspondingly,
$G(E,\bfx_1,\bfx_2)$ and $G^{(2+)}(E,\bfx_1,\bfx_2)$ in the region
where $\bfx_1 \approx \bfx_2$. This means, in particular, that
instead of the original expression for $\Omega$ in
Eq.~(\ref{11a}), $\Omega = \Za/x_1$, one can use its arbitrary
symmetrization with respect to $x_1$ and $x_2$. In our actual
calculations, the following choice of $\Omega$ was employed
\begin{equation}\label{14}
    \Omega  = \frac{2\Za}{x_1+x_2}\,,
\end{equation}
which turned out to be more convenient from the numerical point of
view.

We now use the approximate expression for the Green function
obtained above in order to separate the many-potential term
(\ref{8}) into two parts, one of which contains $G_a^{(2+)}$
instead of $G^{(2+)}$ and is evaluated in a closed form in
configuration space, whereas the remainder is calculated by
summing a rapidly-converging partial-wave series. Bearing in mind
that the partial-wave expansion for the low-energy part of
Eq.~(\ref{8}) is already converging very fast (if the parameter
$\vare_0$ of the integration contour $C_{LH}$ is chosen as
described above), we apply this separation to the high-energy part
only. The many-potential term is thus written as a sum of the
subtraction and the remainder term,
\begin{equation}\label{15}
    \Delta E_{\rm many} = \Delta E_{\rm many}^{\rm
\,sub}+ \Delta E_{\rm many}^{\rm \,remd} \,.
\end{equation}
The subtraction term is obtained from the high-energy part of
Eq.~(\ref{8}) by the substitution $G^{(2+)}\to G^{(2+)}_a$. Its
explicit expression in the Feynman gauge reads
\begin{align} \label{16}
\Delta E_{\rm many}^{\rm \,sub} =& \ \frac{\im\alpha}{2\pi}
     \int_{C_{H}} d\omega \int d\bfx_1\, d \bfx_2\,
\frac{{\rm exp}(\im\,|\omega|\,x_{12})}{x_{12}}\,\,
          \psi^{\dag}_a(\bfx_1)\,
  \nonumber \\ & \times
   \alpha_{\mu}\,
        G^{(2+)}_a(\vare_a-\omega,\bfx_1,\bfx_2)\, \alpha^{\mu}\,
        \psi_a(\bfx_2)
         \,.
\end{align}
The remainder term is obtained from Eq.~(\ref{8}) by applying the
substitution $G^{(2+)}\to G^{(2+)}-G^{(2+)}_a$ in the high-energy
part.

Calculational formulas for the remainder term $\Delta E_{\rm
many}^{\rm \,remd}$ are obtained by obvious modifications of the
corresponding expressions for the many-potential term that can be
found, {\em e.g.}, in Ref.~\cite{yerokhin:99:pra}. In order to
obtain the subtraction term in a form suitable for the numerical
evaluation, one has first to perform the angular part of
integrations over $\bfx_1$, $\bfx_2$ analytically. To do so, we
utilize the fact that both $G^{(2+)}_a$ and the scalar part of the
photon propagator depend on angular variables through $\bfx_{12}$
only. Their product can be written as
\begin{align}\label{17}
    G^{(2+)}_a(\vare_a-\omega,\bfx_1,\bfx_2)\,
   \frac{{\rm exp}(\im\,|\omega|\,x_{12})}{x_{12}}
   \nonumber \\
  = {\cal F}_1\, \im\,\balpha\cdot \bfx_{12} +{\cal F}_2\,\beta+ {\cal F}_3
\,.
\end{align}
Here, ${\cal F}_i \equiv {\cal F}_i(\omega,x_1,x_2,\xi)$ are
scalar functions depending on the radial variables through $x_1$,
$x_2$, and $\xi = \hat{\bfx}_1\cdot\hat{\bfx}_2$ only, where
$\hat{\bfx} = \bfx/x$. Explicit expressions for ${\cal F}_i$ are
immediately obtained from the definition of $G_a^{(2+)}$
(\ref{13}) and the expression for the free Green function
$G^{(0)}$ (\ref{12}). The functions ${\cal F}_i$ can be expanded
over the set of spherical harmonics by
\begin{equation}\label{18}
    {\cal F}_i(\omega,x_1,x_2,\xi) = 4\pi \sum_{l,\,m}
V_l^{(i)}(\omega,x_1,x_2)\,Y_{lm}(\hx_1)\,
     Y_{lm}^*(\hx_2)\,,
\end{equation}
where
\begin{equation}  \label{19}
V_l^{(i)}(\omega,x_1,x_2) = \frac12 \int_{-1}^1 d\xi\, {\cal
F}_i(\omega,x_1,x_2,\xi)\, P_l(\xi)\,
\end{equation}
and $P_l(\xi)$ is a Legendre polynomial.

\begin{widetext}
Substituting Eq.~(\ref{18}) into Eq.~(\ref{16}) and performing
simple angular-momentum algebraic manipulations, we obtain
\begin{align}  \label{20}
\Delta E_{\rm many}^{\rm \,sub} =& \ 2\,\im\, \alpha
     \int_{C_{H}} d\omega \, \int_0^{\infty} dx_1\, dx_2\,
       \int_{-1}^1d\xi\, (x_1x_2)^2
\Bigl\{ {\cal F}_1(\omega,x_1,x_2,\xi)\,g_a(x_1)\,f_a(x_2)
          \left[x_1P_{\overline{l}_a}(\xi)-x_2P_{l_a}(\xi)\right]
   \nonumber \\ &
        +{\cal F}_1(\omega,x_1,x_2,\xi)\,f_a(x_1)\,g_a(x_2)
          \left[x_2P_{\overline{l}_a}(\xi)-x_1P_{l_a}(\xi)\right]
        +2{\cal F}_2(\omega,x_1,x_2,\xi)\,\left[g_a(x_1)\,g_a(x_2)\,P_{l_a}(\xi)
       \right.
   \nonumber \\ &
       \left.
                      - f_a(x_1)\,f_a(x_2)\,P_{\overline{l}_a}(\xi) \right]
        -{\cal F}_3(\omega,x_1,x_2,\xi)\,\left[g_a(x_1)\,g_a(x_2)\,P_{l_a}(\xi)
                      + f_a(x_1)\,f_a(x_2)\,P_{\overline{l}_a}(\xi) \right]
  \Bigr\}\,,
\end{align}
\end{widetext}
where $l_a = |\kappa_a+1/2|-1/2$, $\overline{l}_a = 2j_a-l_a$, and
$g_a(x)$ and $f_a(x)$ are the upper and the lower radial
components of the reference-state wave function $\psi_a(\bfx)$.
The integration over $\omega$ in Eq.~(\ref{20}) can be carried out
analytically in terms of the exponential integral function, as
described in Appendix~\ref{appendix}, leaving a 3-dimensional
integration over the radial variables to be performed numerically.

\section{Numerical evaluation}

The numerical evaluation of the self-energy correction within the
present scheme is in many respects similar to that in the standard
potential-expansion approach. Since the potential-expansion method
is well documented (see, {\em e.g.}, a detailed description in
Ref.~\cite{yerokhin:99:pra}), here we concentrate on novel
features of our evaluation as compared to the standard approach.
They appear in the calculations of ({\em i}) the high-energy part
of the many-potential remainder term $\Delta E_{\rm many}^{\rm
\,remd}$ and ({\em ii}) the many-potential subtraction term
$\Delta E_{\rm many}^{\rm \,sub}$.

The radial integrations over $x_1$ and $x_2$ in the remainder term
$\Delta E_{\rm many}^{\rm \,remd}$ are performed after the change
of variables $(x_1,x_2) \to (r,y)$ \cite{mohr:74:b}:
\begin{equation}\label{21}
 r = {\rm min}(x_1,x_2)/{\rm max}(x_1,x_2)\,,\ \ y =
2\sqrt{1-\vare_a^{\,2}}\,x_2\,.
\end{equation}
Numerical evaluation of the radial integrals is complicated
[specifically, for small values of ${\rm Re}(\omega)$] by the
presence of the function $G^{(0)}(E+\Omega)$ in the integrand. To
explain this, we recall that the analytical behavior of
$G^{(0)}(E+\Omega)$ is governed by the parameter $c^{\,\prime} =
\sqrt{1-(E+\Omega)^2}$. Since $E \equiv \vare_a-\omega =
\vare_a-\varepsilon_0-\im w$ in the high-energy part ($w\in
\mathbb{R}$), the energy argument is
\begin{equation}\label{22}
  E+\Omega =
\vare_a-\varepsilon_0-\im w+ \frac{2\Za}{x_1+x_2}\,.
\end{equation}
For certain values of $x_1$ and $x_2$, ${\rm Re}\,(E+\Omega) = 1$.
When $w$ is small, a fast change of the phase of the square root
$\sqrt{1-(E+\Omega)^2}$ occurs in the vicinity of this point,
which can lead to a numerical instability of the radial
integrations. This problem was handled by breaking the integration
interval at the point where ${\rm Re}\,(E+\Omega) = 1$ and
employing a larger number of integration points in this region.

The numerical evaluation of the subtraction term $\Delta E_{\rm
many}^{\rm \,sub}$ consists of a 3-dimensional integration over
the radial variables, which has a structure of the standard
two-electron integral,
\begin{equation}
    J = \int_0^{\infty}dx_1\,dx_2 \int_{-1}^1 d\xi\,
       \frac{(x_1x_2)^{2}}{x_{12}}\, f(x_1,x_2,\xi)\,,
\end{equation}
where the function $f$ has a finite limit for $x_{12}\to 0$. The
integrable singularity in this expression is removed by employing
the {\em perimetric} coordinates \cite{james:37},
\begin{subequations}
\begin{eqnarray}
 u &=&  x_1+x_2-x_{12}\,, \\
 v &=&  x_1-x_2+x_{12}\,, \\
 w &=& -x_1+x_2+x_{12}\,.
\end{eqnarray}
\end{subequations}
In the new variables, the integral $J$ is
\begin{equation} \label{j1}
    J = \frac14\,\int_0^{\infty}du\,dv\,dw\,
    x_1x_2\,f(x_1,x_2,\xi)\,.
\end{equation}
Performing the integrations in this expression numerically, one
should have in mind that the function $f$ contains a square root,
whose argument changes its sign for certain combinations of the
radial variables, similarly to the case described for the
remainder term $\Delta E_{\rm many}^{\rm \,remd}$. The point at
which the argument of the square root vanishes is
\begin{equation}
\vare_a-\varepsilon_0+ \frac{2\Za}{x_1+x_2} = 1\,.
\end{equation}
This feature was taken into account by breaking the integration
intervals at the singular point and by employing a larger number
of integration points in its vicinity.

\section{Results and discussion}

In Tables \ref{tab:1s}, \ref{tab:2s}, and \ref{tab:2p1} we present
a comparison of two different schemes for the evaluation of the
self-energy correction for the $1s$, $2s$, and $2p_{1/2}$ states.
The labels ``A" and ``B" stand for the subtraction scheme
introduced in this work and for the standard potential-expansion
approach, respectively. The entry ``Free" denotes the sum of the
zero- and one-potential terms (this part is the same in both
methods), ``Subtraction" stands for the many-potential subtraction
term $\Delta E_{\rm many}^{\rm \,sub}$ (absent in the standard
approach), whereas the individual partial-wave expansion
contributions correspond to the many-potential remainder term
$\Delta E_{\rm many}^{\rm \,remd}$ and to the many-potential term
$\Delta E_{\rm many}$ in the ``A" and ``B" schemes, respectively.
The entry ``Behavior" indicates the approximate dependence of the
terms of the partial-wave expansion on $|\kappa|$ in the region of
interest, {\em i.e.}, for $|\kappa| = 10-30$. The numbers in
parentheses represent the uncertainties in the last digit. If no
uncertainties are indicated, numerical values are believed to be
accurate to all digits specified. Our results obtained within the
two approaches are compared with the numerical values by Mohr
\cite{mohr:92:b}.

The comparison of the data listed in the tables demonstrates that
the additional subtraction introduced in this work leads to a
significant improvement of the convergence properties of the
partial-wave expansion in all the cases studied. It also indicates
that the new approach is applicable for the evaluation of the
self-energy correction in the low-$Z$ region, where the standard
potential-expansion approach fails to yield accurate results.

In the low-$Z$ region, one has to deal with numerical
cancellations between individual contributions to the self-energy
correction. The origin of these cancellations are spurious terms
of order $\alpha (\Za)^2\ln \Za$ that appear in the Feynman gauge
when the self-energy correction is separated into the zero, one,
and many-potential terms \cite{snyderman:91} and that have to be
cancelled numerically in order to obtain the physical contribution
to order $\alpha (\Za)^4$. In our approach, the numerical
integrations can be relatively easily performed up to a sufficient
accuracy, so that the numerical cancellations do not pose any
serious problems. Even in the most difficult case, $Z=1$, the
present numerical scheme yields a result with a reasonable
accuracy, $F_{1s}(1\alpha) = 10.316\,85(10)$, which is in a good
agreement with the most precise value by Jentschura {\em et
al.}~\cite{jentschura:01:pra}, $F_{1s}(1\alpha) =
10.316\,793\,650(1)$.

In Table~\ref{tab:total} we present the numerical results for the
self-energy correction in the region that was not previously
tabulated in the literature, $5<Z<10$, and compare our numerical
values for $Z=5$ and $10$ with evaluations by other authors. It is
noteworthy that unlike the previous calculations summarized in
Table~\ref{tab:total}, our evaluation is computationally very
cheap. The time of the calculation for one value of $Z$ is less
than 1h on a modern personal computer. This feature makes the
present approach very promising for extensions to the higher-order
self-energy corrections.

To sum up, we have developed a highly efficient scheme for the
evaluation of the one-loop self-energy correction for an electron
bound in a symmetric local potential (not necessarily the Coulomb
one). The approach presented inherits the attractive features of
the standard potential-expansion method but yields a much better
convergence rate for the resulting partial-wave expansion. As a
result, the applicability of the potential-expansion method is
extended into the region of large values of the parameter
$n^3/(\Za)^2$. We expect that the approach developed will allow
one to significantly improve accuracy of evaluations of the
self-energy correction to the hyperfine splitting and of the
screened self-energy correction in the low-$Z$ region and could be
also applied for higher-order self-energy corrections.

%
\section*{Acknowledgements}

This work was supported by NATO (Grant No.~PST.CLG.979624) and by
RFBR (Grant No.~04-02-17574). V.A.Y. acknowledges also the support
by the "Dynasty" foundation and by INTAS YS grant No.~03-55-1442.

%
%
\begingroup
\squeezetable
\begin{table*}[htb]
\begin{center}
\begin{minipage}{16.0cm}
\caption{Individual contributions to the one-loop self-energy
correction for the $1s$ state, in units of $F(Z\alpha)$. ``A''
denotes the new subtraction scheme, whereas ``B'' indicates the
standard potential-expansion approach. \label{tab:1s}}
\begin{ruledtabular}
\begin{tabular}{l..|..|..}
                  &  \multicolumn{2}{c|}{$Z=5$}&  \multicolumn{2}{c|}{$Z=10$}&  \multicolumn{2}{c}{$Z=92$} \\
                  &  \multicolumn{1}{c}{A} &  \multicolumn{1}{c|}{B}
                                 &  \multicolumn{1}{c}{A} &  \multicolumn{1}{c|}{B}
                                                           &  \multicolumn{1}{c}{A} &  \multicolumn{1}{c}{B} \\
\hline
 Free                                 & -767.72x8\,001      & -767.x728\,0       & -184.0x21\,481      & -184.0x21\,48         & -0.17x1\,545    &   -0.17x1\,545   \\
 Subtraction                          &   30.58x2\,424      &                    &   11.5x27\,613      &                      &  0.29x0\,350    &               \\
 $|\kappa| =     $     1              &  739.69x1\,981      &  759.x830\,8       &  175.7x75\,040      &  183.5x05\,51         &  1.37x1\,144    &    1.63x2\,207   \\
 $\ \ \ \ \ \ \ \ \ \ $2              &    3.43x5\,185      &    8.x855\,9       &    1.2x60\,600      &    3.3x39\,31         & -0.00x1\,514    &    0.01x2\,042   \\
 $\ \ \ \ \ \ \ \ \ \ $3              &    0.22x7\,353      &    2.x299\,5       &    0.0x94\,384      &    0.8x63\,90         &  0.00x1\,728    &    0.00x8\,313   \\
 $\ \ \ \ \ \ \ \ \ \ $4              &    0.02x9\,960      &    1.x028\,4       &    0.0x12\,972      &    0.3x67\,77         &  0.00x0\,469    &    0.00x3\,806   \\
 $\ \ \ \ \ \ \ \ \ \ $5              &    0.00x7\,001      &    0.x568\,2       &    0.0x02\,982      &    0.1x93\,45         &  0.00x0\,155    &    0.00x1\,988   \\
 $\ \ \ \ \ \ \ \ \ \ $6              &    0.00x2\,592      &    0.x352\,0       &    0.0x01\,033      &    0.1x14\,57         &  0.00x0\,062    &    0.00x1\,158   \\
 $\ \ \ \ \ \ \ \ \ \ $7              &    0.00x1\,246      &    0.x234\,7       &    0.0x00\,457      &    0.0x73\,33         &  0.00x0\,029    &    0.00x0\,731   \\
 $\ \ \ \ \ \ \ \ \ \ $8              &    0.00x0\,682      &    0.x164\,8       &    0.0x00\,231      &    0.0x49\,60         &  0.00x0\,015    &    0.00x0\,490   \\
 $\ \ \ \ \ \ \ \ \ \ $9              &    0.00x0\,403      &    0.x120\,2       &    0.0x00\,127      &    0.0x34\,99         &  0.00x0\,008    &    0.00x0\,344   \\
$\ \ \ \ \ \ \ \     $10              &    0.00x0\,250      &    0.x090\,4       &    0.0x00\,073      &    0.0x25\,52         &  0.00x0\,005    &    0.00x0\,251   \\
$\ \ \ \ \ \ \ \     $11              &    0.00x0\,162      &    0.x069\,7       &    0.0x00\,044      &    0.0x19\,12         &  0.00x0\,003    &    0.00x0\,188   \\
$\ \ \ \ \ \ \ \     $12              &    0.00x0\,108      &    0.x054\,8       &    0.0x00\,028      &    0.0x14\,65         &  0.00x0\,002    &    0.00x0\,145   \\
$\ \ \ \ \ \ \ \     $13              &    0.00x0\,074      &    0.x043\,8       &    0.0x00\,018      &    0.0x11\,44         &  0.00x0\,001    &    0.00x0\,114   \\
$\ \ \ \ \ \ \ \     $14              &    0.00x0\,052      &    0.x035\,5       &    0.0x00\,012      &    0.0x09\,08         &  0.00x0\,001    &    0.00x0\,091   \\
$\ \ \ \ \ \ \ \     $15              &    0.00x0\,037      &    0.x029\,2       &    0.0x00\,008      &    0.0x07\,32         &  0.00x0\,001    &    0.00x0\,074   \\
$\sum_{|\kappa|=16}^{35}$             &    0.00x0\,115      &    0.x167\,8       &    0.0x00\,020      &    0.0x38\,82         &  0.00x0\,002    &    0.00x0\,420   \\
$\sum_{|\kappa|=36}^{\infty}$ (extr.) &    0.00x0\,003(2)   &    0.x034(3)       &    0.0x00\,001(1)   &    0.0x07\,2(4)       &  0.00x0\,000    &    0.00x0\,099(3)   \\
 Total                                &    6.25x1\,627(2)   &    6.x252(3)       &    4.6x54\,162(1)   &    4.6x54\,1(4)       &  1.49x0\,916    &    1.49x0\,916(3)   \\
Ref.~\cite{mohr:92:b}                &    \multicolumn{2}{c|}{6.251\,627(8)} &  \multicolumn{2}{c|}{4.654\,162\,2(2)} &  \multicolumn{2}{c}{1.490\,916\,0(3)} \\
 Behavior                             &    30/x|\kappa|^5  &  10x0/|\kappa|^3 & 10x0/|\kappa|^6    &    25x/|\kappa|^3   &  0.5x/|\kappa|^5 &  0.2x5/|\kappa|^3 \\
\end{tabular}
\end{ruledtabular}
\end{minipage}
\end{center}
\end{table*}
\endgroup

%
%
\begingroup
\squeezetable
\begin{table*}[htb]
\begin{center}
\begin{minipage}{16.0cm}
\caption{The same as Table~\ref{tab:1s}, but for the $2s$ state.
\label{tab:2s}}
\begin{ruledtabular}
\begin{tabular}{l..|..|..}
                  &  \multicolumn{2}{c|}{$Z=5$}&  \multicolumn{2}{c|}{$Z=10$}&  \multicolumn{2}{c}{$Z=92$} \\
                  &  \multicolumn{1}{c}{A} &  \multicolumn{1}{c|}{B}
                                 &  \multicolumn{1}{c}{A} &  \multicolumn{1}{c|}{B}
                                                           &  \multicolumn{1}{c}{A} &  \multicolumn{1}{c}{B} \\
\hline
 Free                                 &  -1457.4x18\,809       &-1457.x418\,8         &-356.52x8\,846        &-356.x528\,8           &-1.96x2\,337     &-1.96x2\,337               \\
 Subtraction                          &     31.0x58\,101       &      x               &  11.89x0\,558        &     x                 & 0.27x5\,605  &     x                   \\
 $|\kappa| =     $     1              &   1410.7x15\,203       & 1429.x146\,6         & 339.73x3\,982        & 346.x798\,4           & 3.54x8\,480     & 3.79x6\,632               \\
 $\ \ \ \ \ \ \ \ \ \ $2              &     16.0x99\,356       &   20.x342\,3         &   6.97x9\,535        &   8.x564\,2           & 0.22x8\,968     & 0.20x1\,262               \\
 $\ \ \ \ \ \ \ \ \ \ $3              &      3.6x88\,342       &    5.x475\,2         &   1.72x2\,985        &   2.x395\,0           & 0.07x3\,832     & 0.07x8\,764               \\
 $\ \ \ \ \ \ \ \ \ \ $4              &      1.3x79\,780       &    2.x581\,8         &   0.64x6\,882        &   1.x119\,1           & 0.02x3\,291     & 0.03x4\,477               \\
 $\ \ \ \ \ \ \ \ \ \ $5              &      0.5x63\,792       &    1.x502\,8         &   0.26x3\,506        &   0.x643\,3           & 0.00x7\,670     & 0.01x7\,634               \\
 $\ \ \ \ \ \ \ \ \ \ $6              &      0.2x33\,686       &    0.x979\,1         &   0.10x8\,871        &   0.x414\,0           & 0.00x2\,586     & 0.01x0\,043               \\
 $\ \ \ \ \ \ \ \ \ \ $7              &      0.0x96\,744       &    0.x685\,1         &   0.04x4\,967        &   0.x286\,3           & 0.00x0\,889     & 0.00x6\,184               \\
 $\ \ \ \ \ \ \ \ \ \ $8              &      0.0x39\,919       &    0.x503\,9         &   0.01x8\,535        &   0.x208\,1           & 0.00x0\,312     & 0.00x4\,042               \\
 $\ \ \ \ \ \ \ \ \ \ $9              &      0.0x16\,454       &    0.x384\,5         &   0.00x7\,645        &   0.x157\,1           & 0.00x0\,114     & 0.00x2\,770               \\
$\ \ \ \ \ \ \ \     $10              &      0.0x06\,808       &    0.x301\,9         &   0.00x3\,174        &   0.x122\,0           & 0.00x0\,044     & 0.00x1\,972               \\
$\ \ \ \ \ \ \ \     $11              &      0.0x02\,852       &    0.x242\,5         &   0.00x1\,340        &   0.x096\,9           & 0.00x0\,018     & 0.00x1\,449               \\
$\ \ \ \ \ \ \ \     $12              &      0.0x01\,228       &    0.x198\,5         &   0.00x0\,585        &   0.x078\,5           & 0.00x0\,008     & 0.00x1\,094               \\
$\ \ \ \ \ \ \ \     $13              &      0.0x00\,555       &    0.x165\,0         &   0.00x0\,270        &   0.x064\,5           & 0.00x0\,004     & 0.00x0\,845               \\
$\ \ \ \ \ \ \ \     $14              &      0.0x00\,271       &    0.x139\,0         &   0.00x0\,136        &   0.x053\,8           & 0.00x0\,003     & 0.00x0\,665               \\
$\ \ \ \ \ \ \ \     $15              &      0.0x00\,147       &    0.x118\,4         &   0.00x0\,076        &   0.x045\,3           & 0.00x0\,002     & 0.00x0\,533               \\
$\sum_{|\kappa|=16}^{35}$             &      0.0x00\,400       &    0.x839\,6         &   0.00x0\,207        &   0.x297\,8           & 0.00x0\,005     & 0.00x2\,851               \\
$\sum_{|\kappa|=36}^{\infty}$ (extr.) &      0.0x00\,036(10)   &    0.x35(8)          &   0.00x0\,011(7)     &   0.x082(10)          & 0.00x0\,000     & 0.00x0\,62(3)             \\
 Total                                &      6.4x84\,865(10)   &    6.x54(8)          &   4.89x4\,417(7)     &   4.x898(10)          & 2.19x9\,494     & 2.19x9\,49(3)             \\
 Ref.~\cite{mohr:92:b}                &      \multicolumn{2}{c|}{6.484\,8(2)}         &   \multicolumn{2}{c|}{4.894\,45(6)}          &  \multicolumn{2}{c}{2.199\,493\,8(3)}      \\
 Behavior                             &      5/x|\kappa|^4   &10x0/|\kappa|^{2.5} &10/x|\kappa|^{4.5}  &40x/|\kappa|^{2.5}   &10/x|\kappa|^5 & 2/x|\kappa|^3           \\
\end{tabular}
\end{ruledtabular}
\end{minipage}
\end{center}
\end{table*}
\endgroup


\begingroup
\squeezetable
\begin{table*}[htb]
\begin{center}
\begin{minipage}{16.0cm}
\caption{The same as Table~\ref{tab:1s}, but for the $2p_{1/2}$
state. \label{tab:2p1}}
\begin{ruledtabular}
\begin{tabular}{l..|..|..}
                  &  \multicolumn{2}{c|}{$Z=5$}&  \multicolumn{2}{c|}{$Z=10$}&  \multicolumn{2}{c}{$Z=92$} \\
                  &  \multicolumn{1}{c}{A} &  \multicolumn{1}{c|}{B}
                                 &  \multicolumn{1}{c}{A} &  \multicolumn{1}{c|}{B}
                                                           &  \multicolumn{1}{c}{A} &  \multicolumn{1}{c}{B} \\
\hline
 Free                                 & -1520.7x28\,283     & -1520x.728\,3          &-377.8x53\,426    &-377.x853\,4        &-3.96x6\,890      &-3.96x6\,890      \\
 Subtraction                          &    14.3x76\,901     &      x                 &   6.0x31\,862    &     x              & 0.09x4\,695      &     x          \\
 $|\kappa| =     $     1              &  1481.6x88\,696     &  1483x.524\,8          & 361.0x03\,647    & 361.x809\,2        & 3.88x6\,215      & 3.91x0\,643      \\
 $\ \ \ \ \ \ \ \ \ \ $2              &    18.8x63\,382     &    21x.167\,6          &   8.1x00\,485    &   9.x121\,6        & 0.21x9\,596      & 0.22x1\,754      \\
 $\ \ \ \ \ \ \ \ \ \ $3              &     3.8x16\,241     &     5x.880\,8          &   1.7x40\,929    &   2.x651\,9        & 0.05x8\,070      & 0.07x4\,793      \\
 $\ \ \ \ \ \ \ \ \ \ $4              &     1.1x86\,255     &     2x.832\,7          &   0.5x46\,763    &   1.x267\,1        & 0.01x7\,578      & 0.03x2\,082      \\
 $\ \ \ \ \ \ \ \ \ \ $5              &     0.4x17\,922     &     1x.674\,6          &   0.1x93\,629    &   0.x737\,3        & 0.00x5\,943      & 0.01x6\,268      \\
 $\ \ \ \ \ \ \ \ \ \ $6              &     0.1x55\,297     &     1x.103\,9          &   0.0x72\,407    &   0.x477\,4        & 0.00x2\,214      & 0.00x9\,235      \\
 $\ \ \ \ \ \ \ \ \ \ $7              &     0.0x59\,471     &     0x.779\,4          &   0.0x28\,021    &   0.x330\,9        & 0.00x0\,913      & 0.00x5\,686      \\
 $\ \ \ \ \ \ \ \ \ \ $8              &     0.0x23\,323     &     0x.577\,2          &   0.0x11\,192    &   0.x240\,5        & 0.00x0\,420      & 0.00x3\,724      \\
 $\ \ \ \ \ \ \ \ \ \ $9              &     0.0x09\,401     &     0x.442\,7          &   0.0x04\,654    &   0.x181\,1        & 0.00x0\,215      & 0.00x2\,560      \\
$\ \ \ \ \ \ \ \     $10              &     0.0x03\,945     &     0x.348\,9          &   0.0x02\,053    &   0.x140\,2        & 0.00x0\,122      & 0.00x1\,830      \\
$\ \ \ \ \ \ \ \     $11              &     0.0x01\,764     &     0x.281\,1          &   0.0x00\,985    &   0.x110\,9        & 0.00x0\,074      & 0.00x1\,350      \\
$\ \ \ \ \ \ \ \     $12              &     0.0x00\,866     &     0x.230\,4          &   0.0x00\,528    &   0.x089\,3        & 0.00x0\,048      & 0.00x1\,024      \\
$\ \ \ \ \ \ \ \     $13              &     0.0x00\,479     &     0x.191\,7          &   0.0x00\,318    &   0.x073\,0        & 0.00x0\,033      & 0.00x0\,794      \\
$\ \ \ \ \ \ \ \     $14              &     0.0x00\,300     &     0x.161\,5          &   0.0x00\,212    &   0.x060\,5        & 0.00x0\,023      & 0.00x0\,628      \\
$\ \ \ \ \ \ \ \     $15              &     0.0x00\,209     &     0x.137\,5          &   0.0x00\,153    &   0.x050\,7        & 0.00x0\,016      & 0.00x0\,505      \\
$\sum_{|\kappa|=16}^{35}$             &     0.0x00\,928     &     0x.961\,0          &   0.0x00\,668    &   0.x320\,3        & 0.00x0\,054      & 0.00x2\,747      \\
$\sum_{|\kappa|=36}^{\infty}$ (extr.) &     0.0x00\,13(3)   &     0x.32(8)           &   0.0x00\,08(2)  &   0.x079(10)       & 0.00x0\,002      & 0.00x0\,61(3)    \\
 Total                                &    -0.1x22\,77(3)   &    -0x.11(8)           &  -0.1x14\,84(2)  &  -0.x113(10)       & 0.31x9\,341      & 0.31x9\,34(3)    \\
 Ref.~\cite{mohr:92:b}                &  \multicolumn{2}{c|}{$-0.122\,8(2)$ }     &  \multicolumn{2}{c|}{$-0.114\,83(4)$} & \multicolumn{2}{c}{$0.319\,340\,8(4)$}\\
 Behavior                             &  x2/|\kappa|^{3.5}& 130x/|\kappa|^{2.5}  & 7/x|\kappa|^4   & 200x/|\kappa|^3   & 10/x|\kappa|^5  & 2/x|\kappa|^3    \\
\end{tabular}
\end{ruledtabular}
\end{minipage}
\end{center}
\end{table*}
\endgroup

%
%
\begingroup
\begin{table*}[htb]
\begin{center}
\begin{minipage}{16.0cm}
\caption{One-loop self-energy correction, in terms of $F(\Za)$.
\label{tab:total}}
\begin{ruledtabular}
\begin{tabular}{c....l}
    & \multicolumn{1}{c}{$1s$}
         & \multicolumn{1}{c}{$2s$}
                & \multicolumn{1}{c}{$2p_{1/2}$}
                               & \multicolumn{1}{c}{$2p_{3/2}$}  & \multicolumn{1}{c}{Ref.}      \\
\hline
5   &  6.251\,6x27(2)     &  6.484\,8x65(10)    &  -0.122x\,77(3)     &   0.125x\,64(4)    &                          \\
    &  6.251\,6x27(8)     &  6.484\,8x(2)       &  -0.122x\,8(2)      &   0.125x\,6(1)     &   \cite{mohr:92:b}       \\
    &  6.251\,6x270\,78(1)&  6.484\,8x60\,42(1) &  -0.122x\,774\,94(1)&   0.125x\,623\,30(1) &   \cite{jentschura:01:pra}   \\
    &  6.251\,6x20(3)     &                     &                     &                  &   \cite{indelicato:98}   \\
\hline
6   &  5.817\,8x87(1)     &  6.052\,3x12(10)    &  -0.121x\,43(3)     &   0.126x\,43(4)    &                          \\
7   &  5.458\,0x26\,1(8)  &  5.693\,7x56(10)    &  -0.119x\,95(3)     &   0.127x\,31(3)    &                          \\
8   &  5.152\,0x29\,1(8)  &  5.389\,1x69(9)     &  -0.118x\,35(3)     &   0.128x\,27(3)    &                          \\
9   &  4.887\,0x04\,2(8)  &  5.125\,6x56(8)     &  -0.116x\,65(2)     &   0.129x\,29(4)    &                          \\
\hline
10  &  4.654\,1x62\,4(8)  &  4.894\,4x17(7)     &  -0.114x\,84(2)     &   0.130x\,36(4)    &                          \\
    &  4.654\,1x62\,2(2)  &  4.894\,4x5(6)      &  -0.114x\,83(4)     &   0.130x\,36(2)    &   \cite{mohr:92:b}       \\
    &  4.654\,1x61\,9(1)  &  4.894\,4x44\,4(6)  &  -0.114x\,852(2)    &   0.130x\,350\,7(3)  &   \cite{indelicato:98}
\end{tabular}
\end{ruledtabular}
\end{minipage}
\end{center}
\end{table*}
\endgroup

%
\appendix

\section{Integrals over the virtual photon energy in the
subtraction term} \label{appendix}

The integral over $\omega$ in Eq.~(\ref{20}) can be expressed as a
linear combination of the basic integrals $J_i$,
\begin{equation}
    J_i = \im\int_{C_H}\! d\omega\, f_i(\omega)\, \exp [
    (\im\,|\omega| - c)\,x_{12}]\,,
\end{equation}
where $c = \sqrt{1-(\varepsilon-\omega)^2}$, $C_H =
(\vare_0-\im\infty,\vare_0-\im 0\,]+ [\,\vare_0+\im
0,\vare_0+\im\infty)$, $\vare_0>0 $ is the parameter of the
contour, $\varepsilon$ is either $\vare_a$ or $\vare_a+\Omega$,
and the functions $f_i$ are:
\begin{eqnarray}
  f_1(\omega) &=& 1\,,
\\
  f_2(\omega) &=& x_{12}\,\frac{\varepsilon-\omega}{c}\,,\\
  f_3(\omega) &=& c\,,\\
  f_4(\omega) &=& \varepsilon-\omega\,, \\
  f_5(\omega) &=& x_{12}\,\frac{(\varepsilon-\omega)^2}{c}\,.
\end{eqnarray}

Let us evaluate, {\em e.g.}, the integral $J_1$. Introducing the
new variable $y$ by $\omega = \vare_0 + \im y$ $(\omega = \vare_0
- \im y)$ in the upper (lower) part of the contour, we obtain
\begin{align}
    J_1 =&\ -2\, {\rm Re}\, \exp[\im\, \vare_0\, x_{12}]\,
    \nonumber \\ & \times
      \int_0^{\infty}\!dy\,
      \exp\{-[y+\sqrt{1+(y+\im
      \alpha)^2}]\,x_{12}\}\,,
\end{align}
where $\alpha = \varepsilon-\vare_0$. This integral is evaluated
by introducing the new variable
\begin{equation}
t = y+\sqrt{1+(y+\im \alpha)^2} - \sqrt{1-\alpha^2}\,,
\end{equation}
with the result
\begin{align}
    J_1 =&\ \  -{\rm Re}\, \exp[(\im \vare_0-a)\,x_{12}]
    \nonumber \\ & \times
      \left[ \frac1{x_{12}}+\frac1z -x_{12}\exp(z x_{12})\, E_1(z
      x_{12}) \right] \,,
\end{align}
where $a = \sqrt{1-\alpha^2}$, $z = a+\im \alpha$, and $E_1(z)$ is
the exponential integral function. The results for other basic
integrals are:
\begin{align}
    J_2 =&\ \  -{\rm Im}\, \exp[(\im \vare_0-a)\,x_{12}]
    \nonumber \\ & \times
      \left[ 1-  \frac{x_{12}}{z} +x_{12}^{\,2} \exp(z x_{12})\, E_1(z
      x_{12}) \right] \,,
\end{align}
\begin{widetext}
\begin{equation}
    J_3 = -\frac12\,{\rm Re}\, \exp[(\im \vare_0-a)\,x_{12}]
      \left[ \frac1{x_{12}^{\,2}}+\frac{z}{x_{12}}+\frac1{2\,z^2}
      -\frac{x_{12}}{2\,z} +\left(2
       +\frac{x_{12}^{\,2}}{2}\right) \exp(z x_{12})\, E_1(z
      x_{12}) \right] \,,
\end{equation}
\begin{equation}
    J_4 = -\frac12\,{\rm Im}\, \exp[(\im \vare_0-a)\,x_{12}]
      \left[  \frac1{x_{12}^{\,2}}+\frac{z}{x_{12}}-\frac1{2\,z^2}
      +\frac{x_{12}}{2\,z}
      -\frac{x_{12}^{\,2}}{2}\, \exp(z x_{12})\, E_1(z
      x_{12})\right] \,,
\end{equation}
\begin{eqnarray}
    J_5 &=& -\frac12\,{\rm Re}\, \exp[(\im \vare_0-a)\,x_{12}]
      \left[ -4\im \alpha
      -\frac{1}{x_{12}}-\frac2{z}+z
      -\frac{x_{12}}{2\,z^2}
      +\frac{x_{12}^{\,2}}{2\,z}  \right.
         \left.
      {} + x_{12}\,\left(2-\frac{x_{12}^{\,2}}{2}\right) \exp(z x_{12})\, E_1(z
      x_{12}) \right] \,.
      \nonumber \\
\end{eqnarray}
\end{widetext}
All the expressions for the integrals $J_i$ can readily be
evaluated numerically. A detailed description of an algorithm for
the computation of the exponential integral function of a complex
argument can be found in Ref.~\cite{indelicato:98}.


\end{document}